\documentclass[12pt, a4paper]{article}

\usepackage[left=2.5cm, right=2.5cm, top=2.0cm, bottom=2.0cm]{geometry}
\usepackage{amsmath, amssymb, graphicx}
\usepackage[utf8]{inputenc}
\usepackage[T2A]{fontenc}
\usepackage[english]{babel}
\usepackage{amsthm}
\usepackage{color}
\usepackage{hyperref}
\usepackage{here}
\usepackage{epstopdf}
\usepackage[font=footnotesize,width=.7\textwidth,skip=4mm,labelfont=bf,figurename=Fig.]{caption}
\hypersetup{colorlinks,citecolor=blue,linkcolor=blue}

\newcommand {\tr} {\text{Tr}}
\newcommand {\bra} [1] {\langle#1|}
\newcommand {\ket} [1] {|#1\rangle}
\newcommand {\avg} [1] {\langle#1\rangle}
\newcommand {\proj} [1] {|#1\rangle\!\langle#1|}

\newcommand {\ds} {\displaystyle}
\newcommand {\intl} {\int\limits}

\newcommand {\Acan} {\avg{\hat{A}}_\beta^\text{can}}

\theoremstyle{plain}

\newtheorem{prop}{Proposition}

\theoremstyle{remark}

\title{Does the Eigenstate Thermalization\\Hypothesis Imply Thermalization?}
\author{Oleg Inozemcev, Igor Volovich} 
\date{}

\begin{document}

\maketitle

Steklov Mathematical Institute of Russian Academy of Sciences, Moscow, Russia

\vspace{5mm} E-Mail: inozemcev@mi-ras.ru, volovich@mi-ras.ru

\vspace{6cm}
\begin{center}
	\textbf{Abstract}
\end{center}

\begin{center}
	\parbox{12cm}{Eigenstate thermalization hypothesis (ETH) is discussed. We show that one common formulation of ETH does not necessarily imply thermalization of an observable of isolated many body quantum system. To get thermalization one has to postulate the canonical or microcanonical distribution in the ETH-ansatz. More generally, any other average can be postulated in the generalized ETH-ansatz which leads to a corresponding equilibration condition.
}
\end{center}

\newpage
\section{Introduction}
Recent advancement in the field of thermalization in quantum systems has provided a better understanding of these questions (see overviews \cite{review_Rigol}-\cite{Reimann_Dappr}). Now it is known as the \textit{Eigenstate Thermalization Hypothesis (ETH)}. 

These ideas go back to a seminal von Neumann's paper \cite{Neumann29} (discussed in \cite{Goldst_Neumann}) of 1929 where the main notions of statistical mechanics were re-interpreted in a quantum-mechanical way, the ergodic theorem and the H-theorem were formulated and proved. He also noted that, when discussing thermalization in isolated quantum systems, one should focus on physical observables as opposed to wave functions or density matrices describing the entire system.

The next important step was made in 1991 by Deutsch \cite{Deutsch1}. In his paper based on the random matrix theory (RMT), the relationship between diagonal matrix elements of an observable and microcanonical averages was revealed.

A few years later, Srednicki provided a generalization of the RMT prediction. In 1994 for a gas of hard spheres at high energy and low density, he showed \cite{Sredn1} that each energy eigenstate which satisfies Berry’s conjecture predicts a thermal distribution for the momentum of a single constituent particle. He referred to this remarkable phenomenon as \textit{eigenstate thermalizatio}n. One of well-known formulations of ETH \eqref{SrAnsatz} was given by Srednicki in 1999 \cite{Sredn3}. A slightly different formulation was suggested by Rigol et al. \cite{RigolDunjkoOlshanii}. 

At present there are several different formulations of ETH, see for example discussion in \cite{AnzaGogHub}.

Despite many studies, ETH is still a matter of controversy, see for example \cite{ShirMori}-\cite{ShirMori_Reply}. 
Some current progress on ETH, its applications and related topics can be found in recent papers \cite{RigolSredn}-\cite{MurthySredn}.

The subject of our interest in this paper is the common formulation \eqref{SrAnsatz} of ETH \cite{Sredn3} (hereafter \textit{the Ansatz}). We show that this formulation does not necessarily imply thermalization (in the sense described below) of an observable of isolated many body quantum systems. In the ETH framework, to get such thermalization one has to postulate the canonical or microcanonical distribution in the Ansatz. More generally, any other average can also be postulated in the Ansatz which leads to a corresponding equilibration condition. We hope that this consideration will help to clarify some questions related to formulation of ETH.

In the next section we recall the common formulation of ETH given by Srednicki in \cite{Sredn3} and introduce some notations. In section 3 we consider the Ansatz as a sufficient condition for thermalization and discuss how the Ansatz should be written in this context. In section 4 we discuss how the Ansatz is related to the canonical thermal average. In section 5 we answer the question: Does the Ansatz \eqref{SrAnsatz} really imply thermalization of observables of isolated quantum systems. This consideration suggests the natural formulation of Thermalization Condition (TC) which implies such thermalization (section 6). In section 7, there is a generalization of TC which leads to equilibration with any density matrix.

\vspace{1cm}
\section{Common formulation of ETH}

Throughout this paper we will consider a bounded isolated $N$-body quantum system with $N\gg1$.
Let a Hilbert space $\mathcal{H}$ corresponds to one particle, and $\mathcal{H}_N = \mathcal{H}^{\otimes N}$ corresponds to the entire quantum system.
Denote a Hamiltonian of the system as $\hat{H}$, energy eigenvalues as $E_n$, where
$0 < E_0 \leqslant E_1 \leqslant E_2 \leqslant \ldots$, 
and corresponding eigenstates as $\ket{m}$, so that 
\begin{equation}\label{Schrod}
	\hat{H}\ket{n} = E_n\ket{n}, \hspace{5mm} n=0,1,2,\ldots
\end{equation}
The hermitian operator $\hat{A}$ corresponds to an observable $A$ of this system, and 
\begin{equation}\label{matr_elem}
	A_{mn} := \bra{m}\hat{A}\ket{n}
\end{equation}
are matrix elements of operator $\hat{A}$. 
The pure state $\ket{\psi(t)}$ of the system at any time $t$ is 
\begin{equation}\label{state_t}
	\ket{\psi(t)} = \sum_n c_n \,e^{-iE_n t}\ket{n}, \hspace{10mm} \sum_n |c_n|^2 = 1,
\end{equation}
and $\ket{\psi} := \ket{\psi(0)}$.
We will refer to this system as \textit{quantum system} $\mathfrak{S}$.

Thermodynamic entropy $S(E)$ of continuous energy $E$ is defined by
\begin{equation}\label{entropy}
	e^{S(E)} = E \sum_m \delta_\varepsilon(E-E_m), 
\end{equation}
where $\delta_\varepsilon(x)$ is a Dirac delta function that has been smeared just enough to make $S(E)$ monotonic. 
Suppose that mean energy
\begin{equation}\label{avgE}
	E := \sum_m |c_m|^2E_m
\end{equation}
and entropy of the system are extensive, i.e. as $N\to\infty$
\begin{equation}\label{ES}
	E = \varepsilon N + o(N), \hspace{10mm} 
	S(E) = sN + o(N). 
\end{equation}

In \cite{Sredn3} the following formula for the matrix elements of the observable $A$ in the energy eigenstate basis was suggested
\begin{equation}\label{SrAnsatz}
	A_{mn} = \mathcal{A}(E_{mn})\delta_{mn} + e^{-S(E_{mn})/2}f(E_{mn},\omega)R_{mn},
\end{equation}
where ~$E_{mn} := \frac{1}{2}(E_m+E_n)$, ~$\omega := E_m-E_n$, \\
$\mathcal{A}(E)$ is a smooth function, \\ 
$S(E)$ is the thermodynamic entropy at energy $E$, \\
$f(E,\omega)>0$ is an even function of $\omega$ and a smooth function of its arguments, \\
$R_{mn}\in \mathbb{C}$ is a random variable which varies erratically with $m$, $n$, \\
$\mathbb{E}\left[\text{Re}(R_{mn})\right] = \mathbb{E}\left[\text{Im}(R_{mn})\right] = 0$, $\text{Var}\left[\text{Re}(R_{mn})\right] = \text{Var}\left[\text{Im}(R_{mn})\right] = 1$.

We will refer to the formula \eqref{SrAnsatz} as \textit{the Ansatz}.

According to \cite{Sredn3}, the Ansatz \eqref{SrAnsatz} implies thermalization in many-body quantum systems in the sense that: \\
--- time average of the expectation value of an observable is approximately equal to its \textit{canonical thermal average} at the appropriate temperature; and \\
--- the fluctuations of this expectation value about its time average are small. \\
In this paper thermalization will be regarded in this sense, except for the last section.

This significant statement about thermalization is quite surprising because the Ansatz \eqref{SrAnsatz} has nothing like the canonical distribution, but nevertheless the observable allegedly equilibrates to the canonical thermal average.

Two questions arise: \\
1) Does the Ansatz \eqref{SrAnsatz} really imply thermalization in the above sense? \\
2) How this Ansatz as a sufficient condition for thermalization should be modified?

\vspace{3mm}
For the use in what follows, let us introduce some more notations.
Quantum uncertainty of the mean energy is
\begin{equation}\label{Delta}
	\varDelta^2 := \sum_m |c_m|^2 (E_m-E)^2.
\end{equation}
For any $E>0$ the temperature $T>0$ is defined by 
\begin{equation}\label{}
	\frac{1}{T} \equiv \beta := \frac{dS}{dE}.
\end{equation}
The canonical thermal average of an observable $A$ is
\begin{equation}\label{can_avg}
	\Acan := \frac{\tr\,e^{-\beta H}\!\hat{A}} {\tr\,e^{-\beta H}}.
\end{equation}
The microcanonical thermal average of an observable $A$ at energy $E$ is
\begin{equation}\label{mic_avg}
	\avg{\hat{A}}_E^\text{mic} := \frac{1}{\mathcal{N}_{E,\Delta E}} \sum_{|E-E_n|<\Delta E} A_{nn},
\end{equation}
where $[E\!-\!\Delta E,\: E\!+\!\Delta E]$ is an appropriately chosen energy window (see \cite{RigolDunjkoOlshanii}), 
$\mathcal{N}_{E,\Delta E}$ is the number of energy eigenstates in the energy window.

From \eqref{Schrod}, \eqref{matr_elem} we get a surjective map $E_n \mapsto A_{nn}$.
Let piecewise-linear continuous function $\mathcal{L}(E)$ be linear interpolation for the countable set of points 
\[ (E_0, A_{0,0}),\, (E_1, A_{1,1}),\, (E_2, A_{2,2}),\, \ldots \]
It is obvious that $\forall n\in\mathbb{N}_0 \,:\, A_{nn} = \mathcal{L}(E_n)$.

\vspace{1cm}
\section{The Ansatz as a condition for thermalization}
Let us start from the second question above.
Consider the Ansatz \eqref{SrAnsatz} as a sufficient condition for thermalization.

\vspace{2mm}
\textbf{1. Random variable $R_{mn}$.}

The Ansatz \eqref{SrAnsatz} suffer from mixing of random and deterministic variables.
For a given (non-random) Hamiltonian, the RHS of \eqref{SrAnsatz} is random while the LHS is not. In this sense equality \eqref{SrAnsatz} is not quite correct. 

In \cite{Deutsch1,Deutsch2,Reimann_Dappr} Hamiltonians are taken in the form $\hat{H} = \hat{H}_0 + \hat{V},$ where $\hat{H}_0$ is an “unperturbed” part, $\hat{V}$ is a weak “perturbation”. One can think of $\hat{H}_0$ as describing an ideal gas in a box and $\hat{V}$ as describing two-particle interactions. Instead of adding in these interactions explicitly, $\hat{V}$ is modeled by a random matrix from a certain random matrix ensemble with statistical properties which imitate well the main features of the perturbation $\hat{V}$. That is the many-body Hamiltonian $\hat{H}$ resembles a random matrix. Hence, its eigenstates and matrix elements are random as well.

However, for a deterministic Hamiltonian, the formula \eqref{SrAnsatz} is not correct since there is the deterministic value in the LHS, but the RHS includes the random variable $R_{mn}$.

\vspace{2mm}
\textbf{2. Exponentially small off-diagonal matrix elements.}

From the Ansatz \eqref{SrAnsatz} we see that off-diagonal matrix elements $A_{mn}$ are exponentially small in $N$. But from the proof of Proposition 2 about thermalization (see Appendix B), one can see that off-diagonal matrix elements have to be small but not necessarily exponentially small in $N$.

\vspace{2mm}
\textbf{3. Smooth function $\mathcal{A}(E)$.}

In the Ansatz \eqref{SrAnsatz}, $\mathcal{A}(E)$ is postulated as a smooth function. Then in \cite{Sredn3} there is some argumentation that $\mathcal{A}(E)$ is approximately equal to the canonical thermal average. 
From the proof of Proposition 1 (see Appendix A), one can see that $\mathcal{A}(E)$ may be just a continuous piecewise-linear interpolating function $\mathcal{L}(E)$ for the countable set of points $(E_0, A_{0,0}),\, (E_1, A_{1,1}),\, (E_2, A_{2,2}),\, \ldots$

\vspace{2mm}
Taking this consideration into account, conjectural sufficient condition for thermalization might be written as
\[ A_{mn} = \mathcal{L}(E)\delta_{mn} + \alpha_N, \hspace{5mm}  \alpha_N \xrightarrow[N\to\infty]{} 0. \]
 
\vspace{2mm}
\textbf{4. Quantum uncertainty $\varDelta$.}

In \cite{Sredn3} there is an additional assumption 
\[ \varDelta^2\frac{\mathcal{A}''(E)}{\mathcal{A}(E)} \ll 1, \]
($\varDelta$ is defined in \eqref{Delta}) which is used in the argumentation of thermalization of quantum systems obeying ETH. To avoid this assumption one should postulate a slightly different thing. Namely, separating diagonal matrix elements $A_{nn}$ and off-diagonal matrix elements $A_{mn}$, we get the conjectural thermalization condition:
\begin{align}\label{CTC1}
	\left\{ \begin{aligned}		
	&\sum_n |c_n|^2A_{nn} = \mathcal{L}\Big(\sum_n |c_n|^2E_n\Big) + o(1),  \\
	&A_{mn} = A_{mn}(N) \xrightarrow[N\to\infty]{} 0, \hspace{10mm} m\neq n.
	\end{aligned} \right.
\end{align}
The first equality can be written in short and nice form, so we would have
\begin{align}\label{CTC2}
	\left\{ \begin{aligned}		
	&\bra{\psi}\hat{A}\ket{\psi} = \mathcal{L}(E) + o(1),  \\
	&A_{mn} = A_{mn}(N) \xrightarrow[N\to\infty]{} 0, \hspace{5mm} m\neq n.
	\end{aligned} \right.
\end{align}
But this is not the end of the story.

\vspace{1cm}
\section{The Ansatz and canonical thermal average} 

In \cite{Sredn3} there is a statement (we will call it \textit{Proposition 1}) that the function $\mathcal{A}(E)$ can be related to a standard expression in statistical mechanics: the equilibrium value of A, as given by the canonical thermal average (formula 2.10 in \cite{Sredn3}). From the argumentation presented in \cite{Sredn3}, we see that the precise statement actually should be the following:

\begin{prop}[see \cite{Sredn3}]
	Consider the quantum system $\mathfrak{S}$ (see \eqref{Schrod}--\eqref{state_t}). Then ~$\forall\beta>0 ~~\exists E_\beta>0 ~:~$ 
	\begin{equation}\label{Prop1.1}
		\Acan = \mathcal{A}(E_\beta) + O\big(N^{-1}\big),
	\end{equation}
	where
	\begin{equation}\label{Prop1.2}
		\beta = \frac{\partial S(E)}{\partial E}\bigg|_{E=E_\beta}.
	\end{equation}
\end{prop}

\textbf{Remark 1.} This Proposition states that $\Acan$ can be written for any $\beta>0$ in the form \eqref{Prop1.1} and does not state that this equality holds for any $E>0$.
\vspace{2mm}

As we already mentioned, there is no need to postulate the smooth function $\mathcal{A}(E)$ to get thermalization of quantum systems. With a continuous piecewise-linear interpolating function $\mathcal{L}(E)$ instead of $\mathcal{A}(E)$, we reproduce in more detail a proof (still not rigorous) of this proposition in Appendix A.

\vspace{1cm}
\section{Does the Ansatz imply thermalization?} 

Now we turn to the second question from the Section 2.

Define the expectation value of an observable $A$ as
\[ A_t := \bra{\psi(t)}\hat{A}\ket{\psi(t)} = \sum_{m,n} c_m^* c_n\,e^{i(E_m-E_n)t} A_{mn}, \]
and its infinite time average (Сеsarо mean) as
\begin{equation}\label{overlineA}
	\overline{A} := \lim_{\tau\to\infty} \frac{1}{\tau} \int_0^\tau\!\! A_t\,dt.
\end{equation}

In \cite{Sredn3} (formulas 3.6, 3.7) there is a statement about thermalization of quantum systems obeying ETH  which can be written as:

\textit{ Consider the quantum system $\mathfrak{S}$ (see \eqref{Schrod}--\eqref{state_t}).
If ETH \eqref{SrAnsatz} holds for the observable $A$ then for $N\to\infty$: }
\begin{align}
	&1)~~~ \overline{A} = \Acan + O(\varDelta^2) + O(N^{-1}) + O(e^{-S/2}), \\
	&2)~~~ \overline{(A_t-\overline{A})^2} = O(e^{-S}). 
\end{align}

In short, the argumentation for this statement in \cite{Sredn3} is as follows (accurate to some big O's, $N = \text{const} \gg 1$): 
\begin{equation}\label{SrednEqu} 
\begin{split} 
\overline{A} &= \sum_n |c_n|^2 A_{nn}  
\stackrel{\text{ETH}}{=} \sum_n |c_n|^2\mathcal{A}(E_n) = \\
&= \mathcal{A}\Big(\sum_n |c_n|^2 E_n\Big) \equiv \mathcal{A}(E) = \\
&\stackrel{\text{Prop.1}}{=} \Acan \equiv 
\frac{\sum_n \big[ \exp(-\beta E_n) A_{nn} \big]}{\sum_n \exp(-\beta E_n\big)}.
\end{split} 
\end{equation}

Here we see that Proposition 1 is used to turn from $\mathcal{A}(E)$ to $\Acan$. However, one cannot claim that \eqref{Prop1.1} from Proposition 1 will necessarily hold for $E_\beta=E$ for some $\beta$ (see Remark 1). Hence, it is not proved that the Ansatz \eqref{SrAnsatz} implies thermalization.

Moreover, if equalities \eqref{SrednEqu} hold then we would have
\begin{equation}\label{key}
	\sum_n |c_n|^2 A_{nn} = \sum_n \frac{e^{-\beta E_n}}{Z_\beta} A_{nn}.
\end{equation}
In the LHS there is a weighted arithmetic mean of $A_{nn}$ in a general form. And in the RHS there is also a weighted arithmetic mean but in the special form --- canonical thermal average. 
Although this equality may hold for some $c_n,~ E_n,~ A_{nn},~ n\in\mathbb{N}_0,$ (see Example 1 below), one can easily find such $c_n,~ E_n,~ A_{nn},~ n\in\mathbb{N}_0,$ that the two means will differ significantly.

So, the answer to the question in the title of this section is: no, the Ansatz \eqref{SrAnsatz} does not necessarily imply thermalization.

\vspace{1cm}
\section{Thermalization condition} 

Consideration above shows that the Ansatz \eqref{SrAnsatz} does not necessarily imply thermalization.
Obviously, the same is true for the conjectural thermalization condition \eqref{CTC1}, \eqref{CTC2}
based on the Ansatz \eqref{SrAnsatz}.

If we want to get thermalization in the sense above, we should postulate the canonical thermal average in the ETH-ansatz. Then from \eqref{CTC1} we get the obvious thermalization condition.

\vspace{2mm}
\textbf{Thermalization Condition (TC)}: As $N\to\infty$,
\begin{align}
	&1)~~~ \sum_n |c_n|^2A_{nn} = \Acan + o(1), \label{TC1} \\
	&2)~~~ A_{mn}=A_{mn}(N) \longrightarrow 0, \hspace{5mm} m\neq n. \label{TC2}
\end{align}
Using \eqref{state_t} and \eqref{TC2}, we can write the first equality of the TC in the form
\begin{equation}\label{}
	\bra{\psi}\hat{A}\ket{\psi} = \Acan + o(1).
\end{equation}

It is easy to show that TC imply thermalization of an observable of isolated many body quantum systems.

\begin{prop}[thermalization]
	Consider the quantum system $\mathfrak{S}$ (see \eqref{Schrod}--\eqref{state_t}). If TC holds for the observable $A$ then for $N\to\infty$: \\
	1) the infinite time average of $A_t$ is approximately equal to its equilibrium value \eqref{can_avg} at the appropriate temperature
	\begin{equation}\label{Th2.1}
		\overline{A} = \Acan + o(1);
	\end{equation}
	2) the fluctuations of $A_t$ about $\overline{A}$ are small
	\begin{equation}\label{Th2.2}
		\overline{(A_t-\overline{A})^2} = o(1).
	\end{equation}
\end{prop}

The proof is in Appendix B.

This proof shows that with TC there is no need in assumption \eqref{Delta} on quantum uncertainty.	

\vspace{3mm}
\textbf{Example 1.} An observable of any quantum system with
\begin{equation}\label{c_n}
	c_n = \frac{e^{-\beta E_n/2}}{\sqrt{Z_\beta}}, \hspace{10mm} 
	\ket{\psi} = \sum_n \frac{e^{-\beta E_n/2}}{\sqrt{Z_\beta}} \,\ket{n},
\end{equation}
obviously, satisfies TC. 

\vspace{1cm}
\section{Generalization of TC} 

Note that in \cite{RigolDunjkoOlshanii,review_Rigol} by Rigol et al. ETH is formulated as
\[ \bra{\Psi_\alpha}\hat{A}\ket{\Psi_\alpha} = \avg{\hat{A}}^\text{mic}(E_\alpha), \]
where $\hat{H}\ket{\Psi_\alpha} = E_\alpha\ket{\Psi_\alpha}$. Here we see that microcanonical average is postulated in the ETH-ansatz. Similarly, in \cite{review3} the validity of ETH is investigated through the quantity
\[ I_{\text{ETH}}[\hat{A}] := \max_{\phi_n\in\mathcal{H}_{E,\Lambda}} \left| \bra{\phi_n}\hat{A}\ket{\phi_n} - \avg{\hat{A}}^\text{mic} \right| \xrightarrow[N\to\infty]{} 0. \]

Actually, TC can be written not only with the canonical thermal average $\Acan$ for some $\beta$, but also with the microcanonical average $\avg{\hat{A}}_E^\text{mic}$ at energy $E$, and with any other average $\avg{\hat{A}} = \tr(\rho\hat{A})$, where $\rho$ is an arbitrary density matrix. So, we come to 

\vspace{2mm}
\textbf{Equilibration Condition (EC)}: As $N\to\infty$,
\begin{align}
	&1)\quad \sum_n |c_n|^2A_{nn} = \avg{\hat{A}} + o(1), \hspace{5mm} 
	\avg{\hat{A}} = \tr(\rho\hat{A}) \label{EC1} \\
	&2)\quad A_{mn}=A_{mn}(N) \longrightarrow 0, \hspace{5mm} m\neq n. \label{EC2}
\end{align}

Obviously, as TC implies thermalization, EC implies equilibration. 
Proposition 2 holds with the first statement replaced by the following
\begin{equation}
	\overline{A} = \avg{\hat{A}} + o(1).
\end{equation}

If density matrix commutes with Hamiltonian, i.e. $[\rho,\hat{H}] = 0$, then in the same basis we can write
\[ \hat{H} = \sum_n E_n \proj{n} \]
and
\[ \rho = \sum_n \lambda_n \proj{n}, \qquad \sum_n \lambda_n = 1, \qquad \lambda_n \geqslant 0 ~~ \forall n. \]
From \eqref{EC1} we get
\begin{align*}
	\sum_n |c_n|^2A_{nn} &= \tr(\rho\hat{A}) = \sum_{m,n} \bra{m}\rho\ket{n} \bra{n}\hat{A}\ket{m} = \\ &= \sum_{m,n} \lambda_n \delta_{mn}A_{mn} = \sum_n \lambda_n A_{nn}.
\end{align*}
From this equality we see that in the simplest case $c_n=\sqrt{\lambda_n}$. In addition, 
\[ \dot{\rho} =i[\rho,H] = 0 \]
implies that $\rho$ corresponds to a stationary state, according to general expectations that $\rho(t)$ is close to $\rho^\text{st}$ when $t \gg 1$.

\vspace{3mm}
\textbf{Example 2.} An observable of any quantum system with 
\begin{equation}
	c_n = (\mathcal{N}_{E,\Delta E})^{-1/2}, \hspace{10mm}
	\ket{\psi} = (\mathcal{N}_{E,\Delta E})^{-1/2} \sum_n \ket{n}
\end{equation}
(see \eqref{mic_avg}) satisfies EC. This observable equilibrates to the microcanonical average.

\vspace{1cm}
\section{Conclusion}

Eigenstate thermalization hypothesis in the form of the Ansatz \eqref{SrAnsatz} as a sufficient condition for thermalization is discussed. We have showed that observables of bounded isolated many-body quantum systems satisfying the ETH-ansatz \eqref{SrAnsatz} do not necessarily thermalize in the sense above. 

In the ETH framework, to get such thermalization one, in fact, has to postulate it in the ETH-ansatz. 
Then the thermalization condition \eqref{TC1},\:\eqref{TC2} for bounded isolated many-body quantum systems is readily formulated. It is showed that this condition implies needed thermalization (Proposition 2).

More generally, any other average can also be postulated in the ETH-ansatz which leads to a corresponding equilibration condition \eqref{EC1}, \eqref{EC2}.

A further insight is required to get better understanding of thermalization phenomenon of isolated many-body quantum systems.

\vspace{1cm}
\section{Acknowledgments}
This work was supported by the Russian Science Foundation under the grant \\ \textnumero\,19-11-00320.

\newpage
\section*{Appendix A: Proof of Proposition 1}

For any $\beta>0$ the numerator in \eqref{can_avg} is
\begin{align}
	Z \Acan :=& \tr \left[ e^{-\beta H}A \right] = 
	\sum_m \big[ \exp\left(-\beta E_m\big) A_{mm} \right] = \\
	=& \sum_m \left[ \int e^{-\beta E}\, \delta(E-E_m) \,dE \,A_{mm} \right] = \\
	=& \int e^{-\beta E}\, \sum_m \delta(E-E_m) A_{mm} \,dE,
\end{align}
where $E\in(0,+\infty)$ is the integration variable.
Replace $A_{mm}$ with the interpolating function $\mathcal{L}(E)$
\begin{align}
	Z \Acan &= \int e^{-\beta E}\, \sum_m \delta(E-E_m) \mathcal{L}(E_m)\,dE = \\ 
	&= \int e^{-\beta E}\, \mathcal{L}(E) \sum_m \delta(E-E_m)\,dE. \label{l1}	
\end{align}
	
From \eqref{entropy} we have
\begin{equation}\label{delta}
	\frac{e^{S(E)}}{E} = \sum_m \delta_\varepsilon(E-E_m) \approx \sum_m \delta(E-E_m),
\end{equation}
where $\ds \delta_\varepsilon \xrightarrow[\varepsilon\to0]{} \delta$.
Substitute \eqref{delta} into \eqref{l1}
\begin{align}
	Z \Acan = \int \frac{1}{E} \exp\big[S(E)-\beta E\big] \mathcal{L}(E)\,dE.
\end{align}
	
Similarly, the denominator in \eqref{can_avg} is 
\begin{equation}\label{}
	Z := \tr\,e^{-\beta H} = \int \frac{1}{E} \exp\big[S(E)-\beta E\big]\,dE.
\end{equation}
	
Then \eqref{can_avg} takes the form
\begin{equation}\label{ints}
	\Acan = \frac{\ds \int_0^\infty \frac{1}{E}\:\mathcal{L}(E) \,\exp\!{\big[S(E)-\beta E\big]}\,dE}
	{\ds \int_0^\infty \frac{1}{E} \,\exp\!{\big[S(E)-\beta E\big]}\,dE}.
\end{equation}
	
Now we are going to use Laplace's method (see, for example, \cite{Fedoryuk,Bleis_Hand}) to evaluate the integrals in \eqref{ints}. 
Using \eqref{ES}, we can rewrite the exponent in the last expression as 
\begin{align*}
	S(E)-\beta E = sN - \beta\,\varepsilon N + \alpha_N N = Nh(E).
\end{align*}
Then $\Acan$ takes the form
\begin{equation}\label{}
	\Acan = \frac{\ds \int_0^\infty \frac{1}{E}\:\mathcal{L}(E) \,\exp{\!\big[ Nh(E) \big]}\,dE}
	{\ds \int_0^\infty \frac{1}{E}\,\exp{\!\big[ Nh(E) \big]}\,dE}.
\end{equation}
	
Suppose the function $h(E)$ has a unique global maximum at $\tilde{E}\in(0,\infty)$, i.e.
	\[ h(\tilde{E}) = \max_{E\in(0,\infty)} \big\{h(E)\big\}, \]
where $\tilde{E}$ is such that
\begin{equation}\label{beta(E)}
	\frac{d}{dx}\big(S(E)-\beta E\big)\Big|_{E=\tilde{E}} = 0, \hspace{5mm} \text{or} \hspace{5mm} \beta = S'(\tilde{E}).
\end{equation}
Suppose also that $h(E)\in C(0,\infty)$ is sufficiently smooth in some neighborhood $U_1(\tilde{E})$ of $\tilde{E}$, and $\forall m\in\mathbb{N}_0 \,:\, \tilde{E} \neq E_m$. The last condition implies that $\exists U_2(\tilde{E}) \,:\, \mathcal{L}(E)\in C^\infty\big(U_2(\tilde{E})\big)$. If $\exists k\in\mathbb{N}_0 \,:\, \tilde{E} = E_k$, then interpolation (e.g., Lagrange polynomial) in a neighborhood of $E_k$ can be used.
	
Applying Laplace's method, we find $\forall\beta>0$ and $N\to\infty$
\begin{align*}
	\Acan &= 
	\frac{\ds CN^{-1/2}\frac{1}{\tilde{E}}\:\mathcal{L}(\tilde{E}) \exp{\!\big[Nh(\tilde{E})\big]} + O\Big(N^{-3/2}\exp\big[Nh(\tilde{E})\big]\Big)}
	{\ds CN^{-1/2}\frac{1}{\tilde{E}}\: \exp{\!\big[Nh(\tilde{E})\big]} + O\Big(N^{-3/2}\exp\big[Nh(\tilde{E})\big]\Big)} = \\
	&= \frac{\mathcal{L}(\tilde{E}) + O\big(N^{-1}\big)}
	{1 + O\big(N^{-1}\big)} = 
	\mathcal{L}(\tilde{E}) - O(N^{-2}) + O\big(N^{-1}\big) = \\
	&= \mathcal{L}(\tilde{E}) + O\big(N^{-1}\big), 
\end{align*}
which is the statement of Proposition 1. ~$\blacksquare$

\vspace{1cm}
\section*{Appendix B: Proof of Proposition 2}

	Expectation value $A_t$ can be written as
	\begin{align}
	A_t \equiv \bra{\psi(t)}\hat{A}\ket{\psi(t)} = 
	\sum_{m,n} c_m^* c_n\,e^{i(E_m-E_n)t} A_{mn} = \label{At1} \\ 
	= \sum_m |c_m|^2 A_{mm} + 
	\sum_{k,l\neq k} c_k^* c_l\,e^{i(E_k-E_l)t} A_{kl}. \label{At2}
	\end{align}
	Since $E_m$ are non-degenerate energy eigenvalues then, substituting this $A_t$ \eqref{At2} into the definition \eqref{overlineA} of $\overline{A}$, we get
	\begin{align}
	\overline{A} &= \lim_{\tau\to\infty} \frac{1}{\tau} \intl_0^\tau
	\left( \sum_m |c_m|^2 A_{mm} + \sum_{k,l\neq k} c_k^* c_l\,e^{i(E_k-E_l)t} A_{kl}\right)dt = \nonumber \\ 
	&= \sum_m |c_m|^2 A_{mm}. \label{Ces_mean2}
	\end{align}
	Now we use TC, namely substitute the equality \eqref{TC1} into the last expression:
	\[ \overline{A} = \Acan + o(1), \hspace{5mm} N\to\infty, \]
	which is the first statement of Proposition 2.
	
	Now we turn to the second statement of the Proposition. By definition
	\begin{equation}\label{fluct}
	\overline{(A_t-\overline{A})^2} = \lim_{\tau\to\infty} \frac{1}{\tau} \intl_0^\tau 
	(A_t - \overline{A})^2dt = 
	\lim_{\tau\to\infty} \frac{1}{\tau} \intl_0^\tau A_t^2\,dt - \overline{A}^2.
	\end{equation}
	From \eqref{At1} we get
	\begin{align}
	A_t^2 &= \sum_{k,l,m,n} c_k^* c_l c_m^* c_n \exp[i(E_k-E_l+E_m-E_n)t] A_{kl}A_{mn} = \nonumber \\
	&= \sum_{k,m} |c_k|^2 |c_m|^2 A_{kk}A_{mm} + 	
	\sum_{k,l\neq k} |c_k|^2 |c_l|^2 A_{kl}A_{lk} + (\text{other terms}). \label{At_squar}
	\end{align}
	From \eqref{Ces_mean2} we obtain 
	\begin{equation}\label{Ces_mean_squar}
	\overline{A}^2 = \sum_{k,m} |c_k|^2 |c_m|^2 A_{kk}A_{mm}
	\end{equation}
	Substitute \eqref{At_squar} and \eqref{Ces_mean_squar} into \eqref{fluct}: 
	\begin{align*}
	\overline{(A_t-\overline{A})^2} &= 
	\sum_{k,m} |c_k|^2 |c_m|^2 A_{kk}A_{mm} + 	
	\sum_{k,l\neq k} |c_k|^2 |c_l|^2 A_{kl}A_{lk} - \\
	&-\sum_{k,m} |c_k|^2 |c_m|^2 A_{kk}A_{mm} =
	\sum_{k,l\neq m} |c_k|^2 |c_l|^2 |A_{kl}|^2 \leqslant \\
	&\leqslant \max_{k\neq l}|A_{kl}|^2 \sum_{k,l} |c_k|^2 |c_l|^2 =
	\max_{k\neq l}|A_{kl}|^2. 	
	\end{align*}
	Using TC \eqref{TC2}, we finally get 
	\[ \overline{(A_t-\overline{A})^2} \leqslant \max_{k\neq l} \left\lbrace|A_{kl}(N)|^2\right\rbrace = o(1), \hspace{5mm} N\to\infty, \]
	which is the second statement of Proposition 2. ~$\blacksquare$

\end{document}